\documentclass{llncs}
\usepackage{amsmath,scalerel}
 \usepackage{algorithm}
\usepackage{algorithmic}
\usepackage{tabularx, booktabs}
\usepackage{multirow}
\usepackage{color}
\usepackage{graphics}

\newcolumntype{C}{>{\centering\arraybackslash}p{10ex}}

\graphicspath{{./figures/}}

\begin{document}

\title{Venue Suggestion Using Social-Centric Scores}

\author{Mohammad Aliannejadi and Fabio Crestani}
\institute{Faculty of Informatics, Universit\`a della Svizzera italiana, Lugano, Switzerland\\
\email{\{mohammad.alian.nejadi, fabio.crestani\}@usi.ch}}

\maketitle
\begin{abstract}
User modeling is a very important task for making relevant suggestions of venues to the users. These suggestions are often based on matching the venues' features with the users' preferences, which can be collected from previously visited locations. 
In this paper, we present a set of relevance scores for making personalized suggestions of points of interest. 
These scores model each user by focusing on the different types of information extracted from venues that they have previously visited.
In particular, we focus on scores extracted from social information available on location-based social networks.
Our experiments, conducted on the dataset of the TREC Contextual Suggestion Track, show that social scores are more effective than scores based venues' content.
\end{abstract}

\section{Introduction}
Recent years have witnessed an increasing use of location-based social networks (LBSNs) such as Yelp, TripAdvisor, and Foursquare. These social networks collect valuable information about users' mobility records, which often consist of their check-in data and may also include users' ratings and reviews. 
Therefore, being able to provide personalized suggestions to users plays a key role in satisfying the user needs on such social networks. 
Moreover, LBSNs collect very valuable information from social interactions of users. For instance, the rating history of a user's friends on a social network can be leveraged to improve a recommender system's performance~\cite{DBLP:conf/cikm/RafailidisC16}. Other works have shown that the recommendation can be improved using information from LBSN users who are not in a particular user's friendship network~\cite{yang2015opinions}. 
Also, Foursquare has developed some algorithms to extract informative keywords (called \textit{venue taste keywords}) from users' online reviews. These keywords can be used not only for browsing the reviews more effectively, but also for modeling users. For example, in our previous work~\cite{DBLP:conf/ecir/Aliannejadi17}, we proposed a frequency-based score incorporating venue taste keywords while modeling users.

Recent research has focused on recommending venues using collaborative-filtering technique~\cite{DBLP:conf/recsys/GaoTHL13,8661539}, where the system recommends venues based on users whose preferences are similar to those of the target user (i.e., the user who receives the recommendations). Collaborative-filtering approaches are very effective, but they suffer from the cold-start (i.e., they need to collect enough information about a user for making recommendations) and the data-sparseness problems. Furthermore, these approaches mostly rely on check-in data to learn the preferences of users, and such information is insufficient to get a complete picture of what the user likes or dislikes about a specific venue (e.g., the food, the view). In order to overcome this limitation, we model the users by performing a deeper analysis on users' past ratings as well as their reviews. In addition, following the principle of collaborative filtering, we exploit the reviews from different users with similar preferences. 

In this paper, we present a set of similarity scores for suggesting venues to users, where the users are modeled based on venues' content as well as social information.
Venues' categories are considered as content and online reviews on LBSNs are considered as social information.
Mining social reviews help a system understand the reasons behind a rating: was it for the quality of food, for the good service, for the cozy environment, or for the location? In cases where we lack reviews from some of the users (e.g., they have rated a venue but chose not to review it), we cannot extract opinions, we apply the collaborative filtering principle and we use reviews from other users with similar interests and tastes. Our intuition is that a user's opinion regarding an attraction could be learned based on the opinions of others who expressed the same or similar rating for the same venue. 
To do this we exploit information from multiple sources and combine them to gain better performance. 

This paper extends our previous works~\cite{DBLP:conf/airs/AliannejadiMC16,DBLP:conf/ecir/Aliannejadi17,DBLP:journals/tois/AliannejadiC18} focusing on the social aspects of user modeling. In particular, we have extended the experiments and discussions where we study the impact of using multiple social-centric scores on the performance.
The remainder of the paper is organized as follows. Section~\ref{RelatedWork} reviews related work. Then, we present our methodology in Section~\ref{UserModeling}. Section~\ref{Experiments} describes our experiments. Finally, Section~\ref{Conclusions} is a short conclusion and description of future work.

\section{Related Work}\label{RelatedWork}
Recommender systems try to predict the users' preferences in order to help them find interesting items.
Research on recommender systems was first conducted in the 90s~\cite{DBLP:conf/uai/BreeseHK98}, and since then it has attracted a lot of attention for recommending products in e-commerce websites or information~\cite{DBLP:conf/amt/BahrainianBSD10,DBLP:journals/csur/GiachanouC16} (e.g., news, tweets). 
Recently, due to the availability of the Internet access on mobile devices and based on the fact that users interact with LBSNs more often, researchers have been focusing their interest in analyzing social aspects while recommending venues.

Much work has been carried out in this area based on the core idea that users with similar behavioral history tend to act similarly~\cite{DBLP:journals/cacm/GoldbergNOT92}. This is the underlying idea of collaborative filtering based (CF-based) approaches~\cite{DBLP:conf/recsys/GriesnerAN15,DBLP:journals/tois/YinCSHC14}. 
CF can be divided into two categories: memory-based and model-based. Memory-based approaches consider user rating as a similarity measure between users or items~\cite{DBLP:conf/www/SarwarKKR01}. Model-based approaches, on the other hand, employ techniques like matrix factorization~\cite{koren2008factorization}.
However, CF-based approaches often suffer from data sparsity since there are a lot of available locations, and a single user can visit only a few of them. As a consequence, the user-item matrix of CF becomes very sparse, leading to poor performance in cases where there is no significant association between users and items. Many studies have tried to address the data sparsity problem of CF by incorporating additional information into the model~\cite{DBLP:conf/sigir/YeYLL11,DBLP:conf/sigir/YuanCMSM13}. More specifically, Ye et al.~\cite{DBLP:conf/sigir/YeYLL11} argued that users check-in behavior is affected by the spatial influence of locations and proposed a unified location recommender system incorporating spatial and social influence to address the data sparsity problem. Yin et al.~\cite{DBLP:journals/tois/YinCSHC14} proposed a model that captures user interests as well as local preferences to recommend locations or events to users when they are visiting a new city. 

Yuan et al.~\cite{DBLP:conf/cikm/YuanCS14} proposed to consider both geographical and temporal influences while recommending venues to the users via a geographical-temporal influence-aware graph. They proposed to propagate these influences using a breadth-first strategy.
Ference et al.~\cite{DBLP:conf/cikm/FerenceYL13} took into consideration user preference, geographical proximity, and social influences for venue recommendation. Zhang and Chow~\cite{DBLP:conf/sigir/ZhangC15} exploited geographical, social, and categorical correlations. They modeled the geographical correlation using a kernel estimation method and the categorical correlation by applying the bias of a user on a venue category.  The social check-in frequency or rating was modeled as a power-law distribution to employ the social correlations between users. 
Zhang et al.~\cite{DBLP:journals/tois/ZhangLW16} considered three travel-related constraints (i.e., uncertain traveling time, diversity of the venues, and venue availability) and use them to prune the search space. Griesner et al.~\cite{DBLP:conf/recsys/GriesnerAN15} also proposed an approach integrating temporal and geographic influences into matrix factorization.
In a more recent work, Li et al.~\cite{DBLP:journals/tois/LiJHL17} introduced a fourth-order tensor factorization-based recommendation system considering users' time-varying behavioral trends while capturing their long-term and short-term preferences simultaneously.
Aliannejadi et al.~\cite{DBLP:conf/ecir/Aliannejadi17} proposed a probabilistic mapping approach to determine the most salient information from a venue's content to reduce the dimensionality of data, and extended it to consider the appropriateness of a venue, given a user's context while ranking the venues~\cite{DBLP:conf/sac/AliannejadiMC17,alianSigir17}. 
Yuan et al.~\cite{DBLP:conf/ictai/YuanJGCYA16} addressed the data sparsity problem assuming that users tend to rank higher the venues that are geographically closer to the one that they have already visited. 

Another line of research focuses on enhancing recommendation using users' reviews on LBSNs. When a user writes a review about a venue, there is a wealth of information which reveals the reasons why that particular user is interested in a venue or not. 
Chen et al.~\cite{DBLP:journals/umuai/ChenCW15} state three main reasons for which the reviews can be beneficial for a recommender system: (1) extra information that can be extracted from reviews enables a system to deal with large data sparsity problem; (2) reviews have been proven to be helpful to deal with the cold-start problem; (3) even in cases when the data is dense, they can be used to determine the quality of the ratings or to extract user's contextual information. 
Also, research has shown that venue reviews are effective in determining how similar are two venues~\cite{DBLP:conf/ictir/AliannejadiRC18,DBLP:conf/iir/AliannejadiC18}
Zhang et al.~\cite{DBLP:journals/tist/ZhangDCLZ13} fused virtual ratings derived from online reviews into CF.
Yang and Fang~\cite{DBLP:conf/trec/Yang015} demonstrated how it is possible to get improved recommendations by modeling a user with the reviews of other users' whose tastes are similar to the ones of the target user. In particular, they modeled users by extracting positive and negative reviews to create positive and negative profiles for users and venues. The recommendation is then made by measuring and combining the similarity scores between all pairs of profiles. The effectiveness of online reviews was also shown in more recent works~\cite{alianTREC2016}.

In this paper, we focus on modeling users based on available information on LBSNs. While the available information also includes venues' content (e.g., opening hours), the majority of it is the information left by active users on these social networks. We demonstrate how this type of information helps a recommender system and how a recommender system can leverage it to improve its effectiveness.
\section{Venue Suggestion}\label{UserModeling}
In this section, we first describe the frequency-based scores based on the venues' categories and keywords extracted from Foursquare reviews. Then, we present how to leverage online reviews for venue suggestion.

\subsection{Frequency-based Score}\label{ContextModeling}
We base the frequency-based scores on the assumption that users prefer the type of locations that they like more frequently and rate them positively\footnote{We consider reviews with rating [4, 5] as positive, 3 as neutral, and [1, 2] as negative.}. Therefore, we create positive and negative profiles considering the content of locations in the user's check-in history and calculate the normalized frequencies as they appear in their profile. Then we compute a similarity score between the user's profile and a new location. For simplicity, we only explain how to calculate the frequency-based score using venue keywords. The method can be easily generalized to calculate the score for venue categories.

Let $u$ be a user and $h_u = \{v_1, \dots, v_n\}$ their history of check-ins. Each location has a list of keywords
$C(v_i) = \{c_1, \dots, c_k\}$. We define the user category profile as follows:
\begin{definition}
    \label{df:profile}
    A \textbf{Positive Keyword Profile} is the set of all unique keywords belonging to venues that user $u$ has previously rated positively. A \textbf{Negative Keyword Profile} is defined analogously for venues that are rated negatively.
\end{definition}
Each keyword in the positive/negative keyword profile is assigned with a user-level normalized frequency. We define the user-level normalized frequency for a keyword as follows:
\begin{definition}
    \label{df:freq}
    A \textbf{User-level Normalized Frequency} for an item (e.g., keyword) in a profile (e.g., positive keyword profile) for user $u$
    is defined as: 
    
    \[
       \mathrm{cf}_u^+(c_i) = \frac{\sum_{v_k \in h_u^+}\sum_{c_j \in C(v_k), c_j=c_i} 1}{\sum_{v_k \in h_u}\sum_{c_j \in C(v_k)} 1}~,
    \]
     where $h_u^+$ is the set of locations that $u$ rated positively. We calculate user-level normalized frequency for negative keywords, $\mathrm{cf}_u^-$, analogously. 
    
\end{definition}


\subsubsection{Foursquare Taste Keywords.}
Foursquare automatically extracts a list of keywords, also known as ``tastes'' to better describe a venue. These keywords are extracted from online reviews of users who visit a venue. As an example, ``Central Park'' in ``New York City'' is described by these taste terms: \textit{picnics, biking, trails, park, scenic views,} etc.  Such keywords are very informative, since they often express characteristics of a venue, and they can be considered as a complementary source of information for venue categories. 

Table \ref{tb:keywords} shows all taste keywords and categories for a sample restaurant on Foursquare. As we can see, the taste keywords represent much more details about the venue compared to categories. The average number of taste keywords for venues (8.73) is much higher than the average number of categories for venues (2.8). It suggests that these keywords could describe a venue in more details compared to categories. 

We create positive and negative keyword profiles for each user based on Definitions \ref{df:profile} and \ref{df:freq}. Given a user $u$ and candidate venue $v$, the frequency-based similarity score based on venue keywords, $S_{key}(u,v)$, is calculated as follows:
\begin{equation}
    \label{eq:cat}
    S_{key}(u,v) = \sum_{c_i \in C(v)}\text{cf}_u^+(c_i) - \text{cf}_u^-(c_i)~.
\end{equation}

\subsubsection{Venue Categories.} 
Here we aim to exploit the categories of the venues a user liked in the past. Such information represents an important information that can be used to infer what kind of places a user may enjoy visiting. In some cases, categories are the only source of information. For example, a venue that has not received many online reviews.
We adopt the same frequency-based approach as we did for venue taste keywords. Thus, we create positive and negative category profiles for user considering venue categories, based on Definitions \ref{df:profile} and \ref{df:freq}. Then, we compute the category similarity score, $S_{cat}(u,v)$, as we did for the keyword-based score (see Equation (\ref{eq:cat})).

\begin{table}[t]
    \centering
    \caption{A sample of taste keywords and categories for a restaurant}
    \begin{tabular}{l@{\quad}l}
        \toprule
         \textbf{Taste keywords} & pizza, lively, cozy, good for dates, authentic, casual, pasta, desserts\\
                        &  good for a late night, family-friendly, good for groups, ravioli, \\ 
                        & lasagna, salads, wine, vodka, tagliatelle, cocktails, bruschetta \\ 
        \midrule
         \textbf{Categories} & pizza place, italian restaurant \\ 
         \bottomrule
    \end{tabular}
    \label{tb:keywords}
\end{table}

\subsection{Review-Based Score}\label{sec:rev}
Modeling a user only on locations' content is general and does not determine why the user enjoyed or disliked a venue.
The content of locations is often used to infer ``which type'' of venues, a user likes. On the other hand, reviews express the reasons for users' ratings. Since there could be a lack of explicit reviews from the user, we tackle this sparsity problem using reviews of other users who gave a similar rating to the location. In particular, we calculate the review-based score using a binary classifier. 

We model this problem as binary classification since a user, before visiting a new city or location, would get a positive or negative impression of the location after reading the online reviews of other users. We assume that a user would measure the characteristics of a location according to their expectations and interests. These characteristics are mainly inferred from the existing online reviews of other users.
The user would be convinced to visit a particular location if the reviews satisfy their expectations up to a certain point.
An alternative to binary classification would be a regression model, however, we assume that users behave like a binary classifier when they read online reviews in order to make a decision on whether to visit a venue or not. For example, assume a user reads a few positive and negative online reviews about a venue and measures how similar the mentioned qualities are to their expectations. Finally, depending on the balance between the positive remarks and the negative ones, they make a binary decision (i.e., whether to go or not). We see this behavioral pattern similar to that of a binary classifier: it learns from the positive and negative samples and compares the learned parameters with a test sample and assigns its label accordingly.
Furthermore, due to data sparsity, grouping ratings as positive and negative aids us to model users more effectively.

For each user, we train a binary classifier using the reviews from the locations in a user's check-in history. 
The positive classification training samples for user $u$ are positive reviews of locations that were liked by $u$. Likewise, the negative reviews of locations that $u$ disliked constitute the negative training samples.
We decided to ignore the negative reviews of liked locations and positive reviews of disliked locations since they are not supposed to contain any useful information.

After removing the stop words, we consider the TF-IDF score of terms in reviews as features. We trained many classifiers but linear SVM outperformed all other models. Therefore, we choose linear SVM and consider the value of the its decision function as the review-based score and refer to it as $S_{rev}(u, v)$. The decision function gives us an idea on how relevant a location is to a user profile. We used the scikit-learn\footnote{\url{http://scikit-learn.org/}} implementation of SVM with default parameters (i.e., penalization: $l2$-norm, loss function: squared hinge, c=1.0).

\subsection{Location Ranking}\label{sec:ranking}
After defining the mentioned relevance scores, here we explain how we combine them.
Given a user and a list of candidate locations, we calculate the mentioned scores for each location and combine them to create a ranked list of locations. We adopt several learning to rank\footnote{We use RankLib implementation of learning to rank: \url{https://sourceforge.net/p/lemur/wiki/RankLib/}} techniques to rank the candidate locations since they have proven to be effective for similar tasks~\cite{DBLP:journals/ftir/Liu09}. In particular, we examine the following learning to rank techniques: AdaRank, Coordinate Ascent (aka. CAscent), RankBoost, MART, $\lambda$-MART, RandomForest, RankNet, and ListNet. We study the performance of different five models using different combinations of the scores as follows:

\begin{itemize}
    \item \textbf{LTR-All}: This model consists of all proposed relevance scores: $S_{cat}$ (from both Yelp and Foursquare), $S_{rev}$, and $S_{key}$.
    \item \textbf{LTR-S}: It consists only of the social-centric scores: $S_{rev}$ and $S_{key}$.
    \item \textbf{LTR-C}: It includes only of non social scores: $S_{cat}$ (from both Yelp and Foursquare).
    \item \textbf{LTR-Y}: We only include the scores calculated using Yelp: $S_{cat}$ (only from Yelp) and $S_{rev}$.
    \item \textbf{LTR-F}: Information from Foursquare is only considered for this model: $S_{cat}$ (only from Foursquare) and $S_{key}$.
\end{itemize}

\section{Experiments}\label{Experiments}
This section describes the dataset, the experimental setup for assessing the performance of our methodology, and the experimental results. 

\subsection{Experimental Setup.}\label{ExperimentSetup}
\subsubsection{Dataset.}
Our experiments were conducted on the collection provided by the Text REtrieval Conference (TREC) for the Batch Experiments of the 2015 Contextual Suggestion Track\footnote{\url{https://sites.google.com/site/treccontext/trec-2015}}. 
This track was originally introduced by the National Institute of Standards and Technology (NIST) in 2012 to provide a common evaluation framework for participants that are interested in dealing with the challenging problem of contextual suggestions and venue recommendation. 
In short, given a set of example places as user's preferences (profile) and contextual information (e.g., the \emph{city} where the venues should be recommended), the task consists in returning a ranked list of $30$ candidate places which match the user's profile. 
The ratings range between 0 (very uninterested) and 4 (very interested).
The collection, provided by TREC, consists of a total $9K$ distinct venues and $211$ users. For each user, the contextual information plus a history of $60$ previously rated attractions are provided. Additionally, for our experiments, we used the additional crawled information released by~\cite{alianSigir17collection}. 

\subsubsection{Evaluation Metrics.}
We use the official evaluation metrics of TREC for this task which are P@5 (Precision at 5), nDCG@5 (Normalized Discounted Cumulative Gain at 5), and MRR (Mean Reciprocal Rank). In order to find the optimum setting of learning to rank techniques, we conducted a 5-fold cross validation with respect to nDCG@5. We determine the statistically significant differences using the two-tailed paired t-test at a $95\%$ confidence interval ($p < 0.05$).

\subsubsection{Compared Methods.}
We compare our proposed method with state-of-the-art context-aware and social-based venue recommendation methods.

\begin{itemize}
    \item \textit{LinearCatRev}~\cite{alian2015} is the best performing model of TREC 2015. It extracts information from different LBSNs and uses it to calculate category-based and review-based scores. Then, it combines the scores using linear interpolation. We choose this baseline for two reasons, firstly because it is the best performing system of TREC 2015, and secondly because it also uses scores derived from different LBSNs.

    \item \textit{GeoSoCa} exploits geographical, social, and categorical correlations for venue recommendation~\cite{DBLP:conf/sigir/ZhangC15}. GeoSoCa models the geographical correlation using a kernel estimation method with an adaptive bandwidth determining a personalized check-in distribution. It models the categorical correlation by applying the bias of a user on a venue category to weigh the popularity of a venue in the corresponding category modeling the weighted popularity as a power-law distribution. It models the social ratings as a power-law distribution employing the social correlations between users.

    \item \textit{n-Dimensional Tensor Factorization (nDTF)}~\cite{DBLP:conf/recsys/KaratzoglouABO10} generalizes matrix factorization to allow for integrating multiple contextual features into the model. Regarding the features, we included two types of features: (1) venue-based: category, keywords, average rating on Yelp, and the number of ratings on Yelp (as an indicator of its popularity); (2) user-based: age group and gender.
\end{itemize}

\subsection{Results and Discussions}\label{ExperimentResults}
In this section, we present a set of experiments in order to demonstrate the effectiveness of our approach. Then, we study the effect of social features on the performance.

\subsubsection{Performance Evaluation Against Compared Methods.}
Table \ref{tb:preliminarily_results_2015} demonstrates the performance of our approach against the compared methods. We chose to report the results obtained by RankNet because it exhibited the best performance among all other learning to rank techniques (see Table~\ref{tb:ltr2015}). Table \ref{tb:preliminarily_results_2015} shows that LTR-S outperforms the competitors with respect to the three evaluation metrics. This shows that using social-centric features can effectively model users on LBSNs leading to higher recommendation performance. Note that LTR-S also outperforms LTR-All which consists of both social- and content-based scores, indicating that category scores are not as effective as social scores. This is also evident in the results obtained by LTR-C, where only category scores are included in the model and the results are much lower than of LinearCatRev. Table~\ref{tb:preliminarily_results_2015} also illustrates the performance of our model when using the scores obtained from only one source of information. In particular, LTR-Y and LTR-F are trained using the scores computed only on Yelp and Foursquare data, respectively. As we can see, they both perform worse than LinearCatRev, suggesting that combining cross-platform social information is critical while recommending venues to users. Finally, we see that GeoSoCa and nDTF exhibit the worst performance among all compared methods. This happens mainly because these methods rely on user-venue check-in associations among the training and test sets. In other words, there should be enough common venues appearing in both the training and test sets, otherwise, they fail to recommend unseen venues. Hence, they suffer from the high level of the sparsity of the dataset. In fact, the intersection of venues in the training and test sets is 771 (out of 8,794).

To train the review-based classifier, we used various classifiers such as Na{\"i}ve Bayes and k-NN; however, the SVM classifier exhibited a better performance by a large margin. The SVM classifier is a better fit for this problem since it is more suitable for text classification, which is a linear problem with weighted high dimensional feature vectors. Also, we observed a significant difference between the number of positive reviews and negative reviews per location. Generally, locations receive more positive reviews than negative reviews and, in our case, this results in an unbalanced training set. Most of the classification algorithms fail to deal with the problem of unbalanced data. This is mainly due to the fact that those classifiers try to minimize an overall error rate. Therefore, given an unbalanced training set, the classifier is usually trained in favor of the dominant class to minimize the overall error rate. However, SVM does not suffer from this, since it does not try to directly minimize the error rate but instead tries to separate the two classes using a hyperplane maximizing the margin. This makes SVM more intolerant of the relative size of each class. Another advantage of linear SVM is that the execution time is very low and there are very few parameters to tune.

\begin{table}[t]
\centering
\caption{Performance evaluation on TREC 2015. Bold values denote the best scores and the superscript * denotes significant differences compared to LinearCatRev. $\Delta$ values ($\%$) express the relative difference, compared to LinearCatRev.}
\label{tb:preliminarily_results_2015}
\begin{tabular}{l@{\quad}l@{\quad}r@{\quad}l@{\quad}r@{\quad}l@{\quad}r}
\toprule
 & \multicolumn{1}{l}{P@5} & \multicolumn{1}{c}{$\Delta(\%)$} & \multicolumn{1}{l}{nDCG@5} & \multicolumn{1}{c}{$\Delta(\%)$} & \multicolumn{1}{l}{MRR} & \multicolumn{1}{c}{$\Delta(\%)$} \\
\midrule
LinearCatRev & 0.5858  & \multicolumn{1}{c}{{-}} & 0.6055 & \multicolumn{1}{c}{{-}} & 0.7404 & \multicolumn{1}{c}{{-}} \\
GeoSoCa & 0.5147* & $-$12.14 & 0.5404* & $-$10.75 & 0.6918* & $-$6.56\\
nDTF & 0.5232* & $-$10.96 & 0.5351* & $-$11.63 & 0.6707* & $-$9.41 \\
LTR-All & $0.5913$ & $0.94$ & $0.6087$ & $0.53$ & $0.7411$ & $0.10$   \\ 
LTR-S & $\mathbf{0.6038}$* & $3.07$ & $\mathbf{0.6235}$* & $2.98$ & $\mathbf{0.7419}$ & $0.21$  \\ 
LTR-C & $0.5376$* & $-8.22$ & $0.5408$* & $-10.69$ & $0.6643$* & $-10.28$  \\ 
LTR-Y & $0.5323$* & $-9.13$ & $0.5334$* & $-11.91$ & $0.6500$* & $-12.20$   \\ 
LTR-F & $0.5558$* & $-5.11$ & $0.5784$* & $-4.47$ & $0.7261$ & $-1.93$   \\ 
\bottomrule
\end{tabular}
\end{table}

\subsubsection{Impact of Different Learning to Rank Techniques.} 
In this experiment, we aim to show how the recommendation effectiveness is affected by applying different learning to rank techniques to combine the scores. Table~\ref{tb:ltr2015} reports nDCG@5 applying different learning to rank techniques for TREC 2015. We report the performance for LTR-All, LTR-S, LTR-C, LTR-Y, and LTR-F. As we can see, RankNet outperforms other learning to rank techniques when using only social-centric features (LTR-S). It is worth noting that RankNet and ListNet are both based on artificial neural networks, and they perform best considering most of the models. As we can observe, applying different learning to rank techniques can potentially have a big impact on recommendation results. Therefore, it is critical to apply the best technique according to the scores.

\begin{table}[t]
\centering
\caption{Effect on nDCG@5 for different learning to rank techniques on TREC 2015. Bold values denote the best scores per model and the superscript * denotes significant differences compared to LinearCatRev. $\Delta$ values ($\%$) express the relative difference, compared to LinearCatRev (nDCG@5 = 0.6055).}
\vspace{-0.1cm}
\resizebox{12cm}{!}{%
\begin{tabular}{llrl@{\quad}lrl@{\quad}lrl@{\quad}lrl@{\quad}lr}
\toprule
 & \multicolumn{1}{l}{LTR-All} & \multicolumn{1}{c}{$\Delta$} && \multicolumn{1}{l}{LTR-S} & \multicolumn{1}{c}{$\Delta$} && \multicolumn{1}{l}{LTR-C} & \multicolumn{1}{c}{$\Delta$} && \multicolumn{1}{l}{LTR-Y} & \multicolumn{1}{c}{$\Delta$} && \multicolumn{1}{l}{LTR-F} & \multicolumn{1}{c}{$\Delta$}\\
\cmidrule{2-3} \cmidrule{5-6} \cmidrule{8-9} \cmidrule{11-12} \cmidrule{14-15}
MART & $0.5899$* & $-2.57$ && $0.5995$ & $-1.00$ && $0.5575$* & $-7.93$ && $0.6023$ & $-0.53$ && $0.5691$* & $-6.01$   \\
RankNet & $0.6087$ & $0.53$ && $\mathbf{0.6235}$* & $2.98$ && $0.5408$* & $-10.69$ && $0.5334$* & $-11.91$ && $0.5784$* & $-4.47$   \\
RankBoost & $0.5924$* & $-2.17$& & $0.5980$ & $-1.23$ && $0.5573$* & $-7.96$ && $0.5891$* & $-2.70$ && $0.5529$* & $-8.69$   \\
AdaRank & $0.6074$ & $0.32$ && $0.6180$ & $2.06$ && $0.5762$* & $-4.84$ && $0.6009$ & $-0.76$ && $0.5735$* & $-5.28$   \\ 
\small{CAscent} & $\mathbf{0.6089}$ & $0.57$ && $0.6160$ & $1.74$ && $\mathbf{0.5763}$* & $-4.82$ && $0.6037$ & $-0.30$ && $0.5768$* & $-4.73$   \\ 
\small{$\lambda$-MART} & $0.6065$ & $0.17$& & $0.6134$ & $1.31$ && $0.5645$* & $-6.77$ && $0.5987$ & $-1.12$ && $0.5724$* & $-5.47$   \\
ListNet & $0.6068$ & $0.21$ && $0.6198$ & $2.36$ && $0.5762$* & $-4.84$ && $\mathbf{0.6066}$ & $0.18$ && $\mathbf{0.5787}$* & $-4.42$   \\ 
\bottomrule
\end{tabular}
}
\label{tb:ltr2015}
\end{table}

\subsubsection{Impact of Number of Reviews.}
\begin{figure}[t]
    \centering
    \begin{tabular}{cc}
         \includegraphics[width=.48\textwidth]{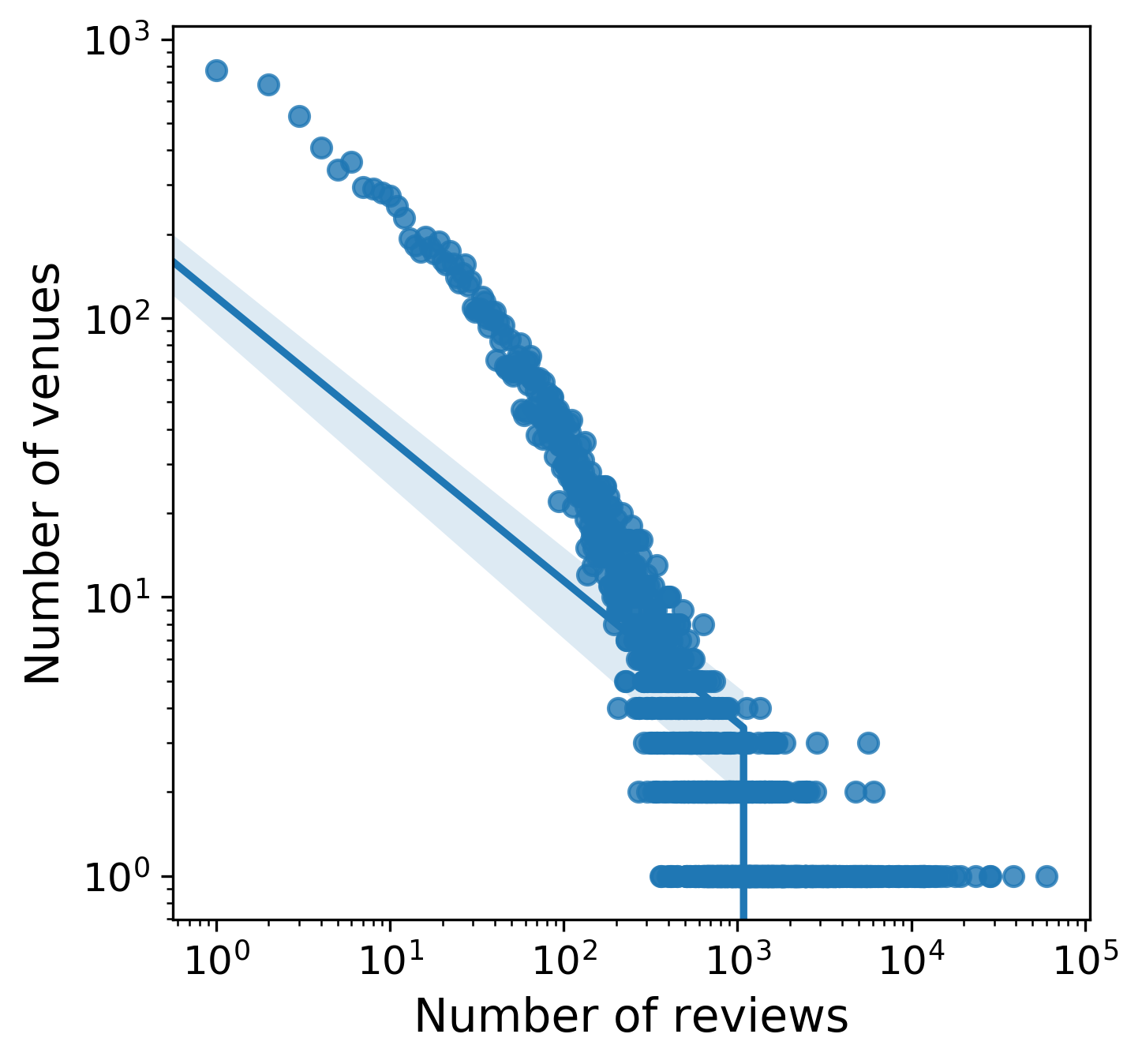} & \includegraphics[width=.48\textwidth]{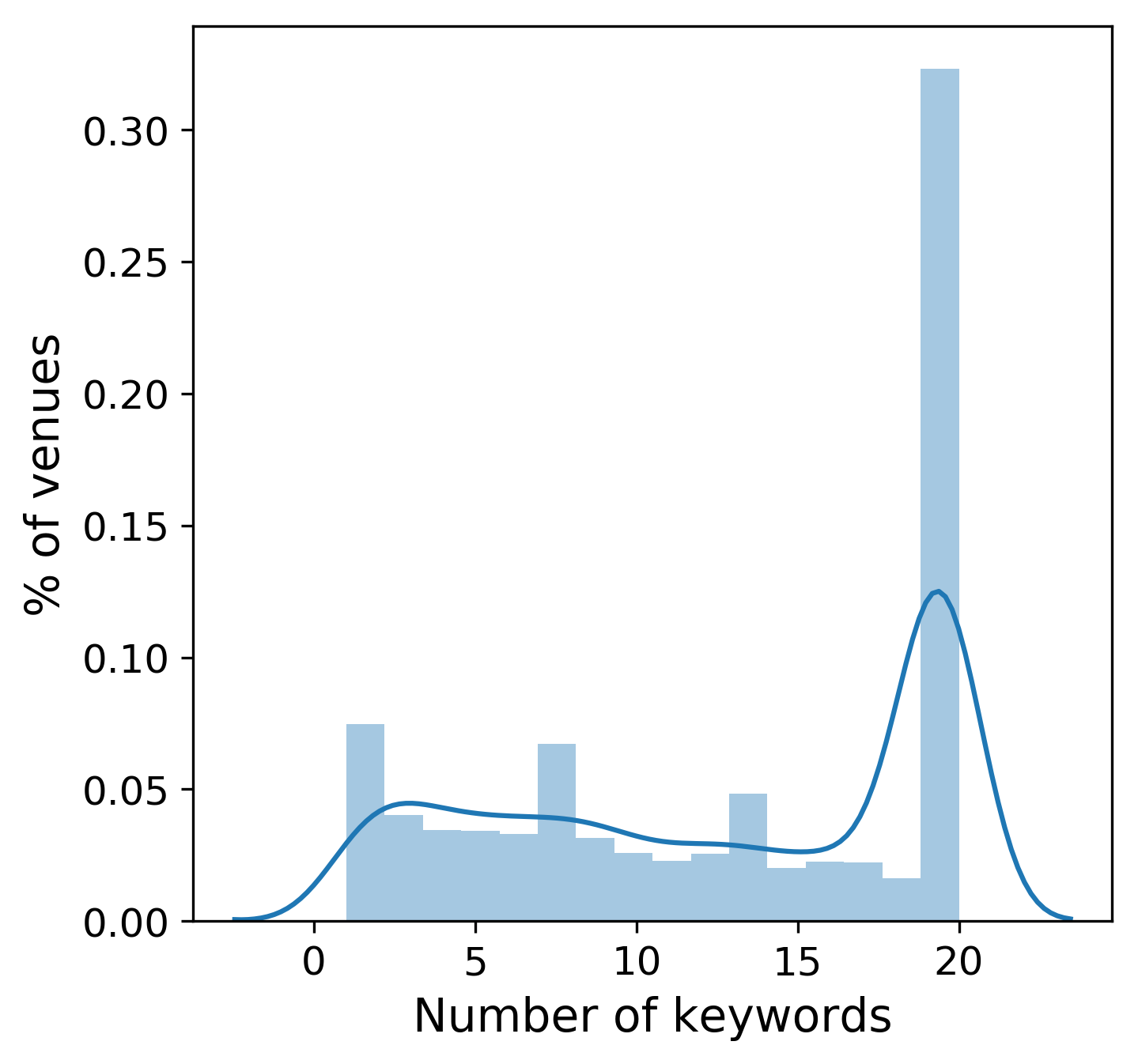} \\
         (a) Reviews & (b) Keywords
    \end{tabular}
    \caption{Distribution of the number reviews and keywords per venue.}
    \label{fig:revcount}
\end{figure}

Here we show how the recommendation effectiveness is affected by the number of online reviews used to compute the review-based score. Users leave a massive number of reviews about venues on LBSNs, making it very difficult for a system to consider all the reviews while modeling users. Figure~\ref{fig:revcount}a illustrates the distribution of reviews per venue, showing that a considerable number venues receive many reviews. Therefore, it is crucial to study the impact of the number of reviews on the performance of our model. 

\begin{figure}[t]
    \centering
    \includegraphics[width=0.8\textwidth]{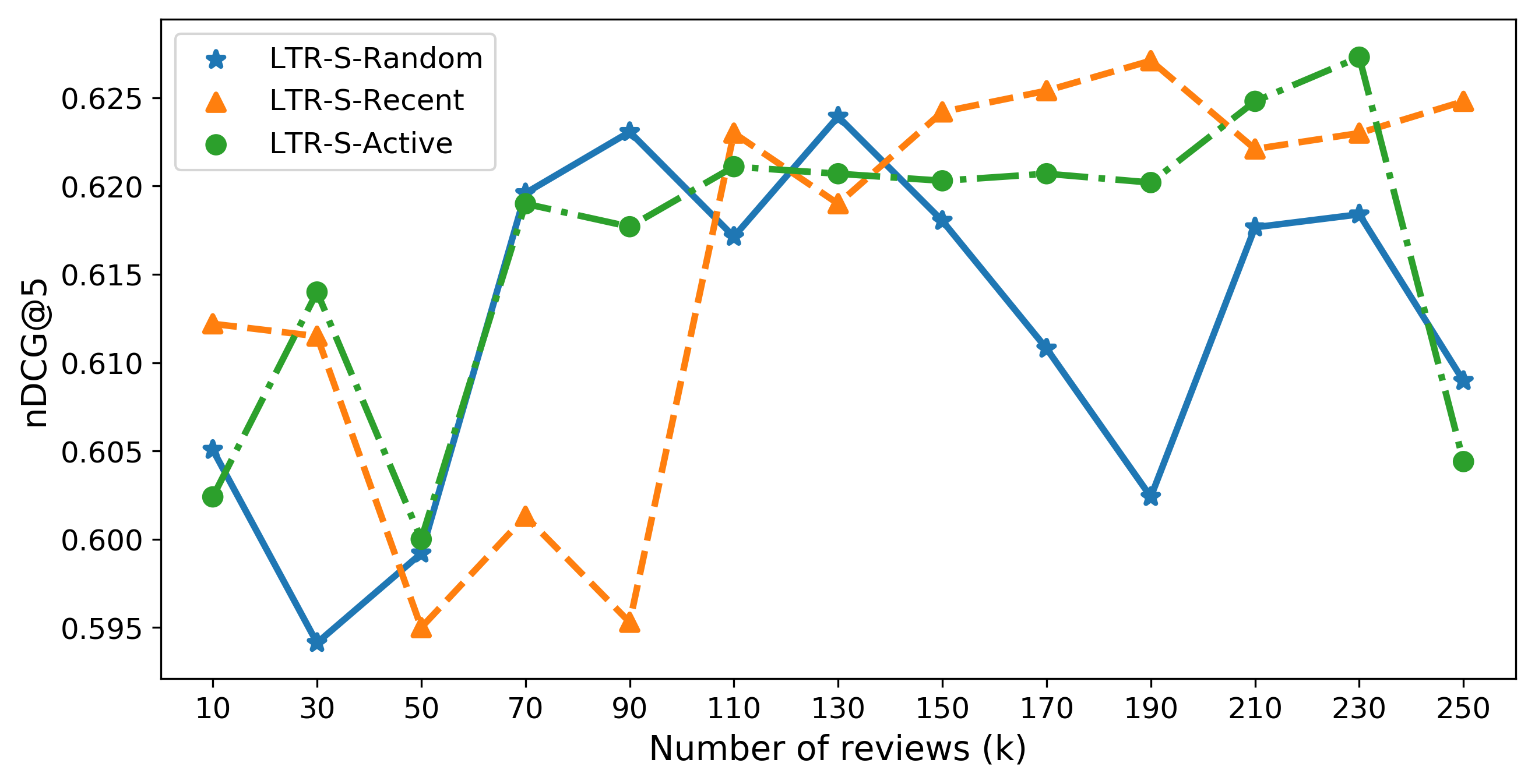}
    \caption{Performance of LTR-S in terms of nDCG@5 using different number of reviews ($k$).}
    \label{fig:rev}
\end{figure}

Figure~\ref{fig:rev} shows the performance of LTR-S as we change the number of reviews while building user profiles. We follow three criteria as we vary the number of reviews:
\begin{itemize}
    \item \textit{LTR-S-Random} selects $k$ reviews per venue randomly. To prevent random bias, we ran this model 5 times and report the average performance. 
    \item \textit{LTR-S-Recent} includes the $k$ most recent reviews in the user profile. Here, we are interested in exploring the temporal effect of reviews.
    \item \textit{LTR-S-Active} builds the review profiles considering the reviews from top $k$ active users. A user activity is measured by the total number of reviews that they have written on Yelp. Here, we are interested in finding out if the users level of activity can be used to determine the credibility of their reviews.
\end{itemize}
 
As we can see, results are comparable to LTR-S when we use only 230 reviews, showing that the model converges after a certain number of reviews. Moreover, using more reviews can potentially have a negative impact, because the model will be biased towards the venues that have a higher number of reviews (i.e., more popular venues). The results of LTR-S-Random exhibit the least consistency as we increase $k$, showing that a random selection of reviews is not as effective as other criteria. We see that both LTR-S-Recent and LTR-S-Active show less consistency with lower $k$'s, but improve as $k$ grows. Specifically, LTR-S-Recent achieves its best performance with $k=190$ (nDCG@5 = 0.6271) and LTR-S-Active with $k=230$ (nDCG@5 = 0.6273), both outperforming LTR-S. This indicates that pruning reviews based on time and user activity improves not only the system's efficiency but also its effectiveness.

\subsubsection{Impact of Number of Keywords.}
In this experiment we study how the recommendation effectiveness is affected by the number of venue taste keywords in user profiles. As discussed in~\cite{DBLP:conf/ecir/Aliannejadi17}, venue taste keywords are very sparse because they are automatically extracted from user reviews and contain various sentimental tags. Moreover, as we can see in Figure~\ref{fig:revcount}b, venue profiles on Foursquare are featured with many keywords and it is crucial to reduce the dimensionality of keywords such that less important keywords are removed from the profiles. We follow three criteria as we vary the number of keywords in the profiles:
\begin{itemize}
    \item \textit{LTR-SKey-VRand} randomly selects $k$ keywords for each venue and creates user profiles using those keywords. Note that since the maximum number of keywords per venue is 20, we vary $k$ from 0 to 20.
    \item \textit{LTR-SKey-URand} creates the user profiles using the full list of keywords but considers only $k$ randomly selected keywords from the user's profile, when computing the relevance score. We vary $k$ from 0 to 300.
    \item \textit{LTR-SKey-UPop} creates the user profiles using the full list of keywords but computes the relevance scores using only $k$ keywords with highest frequencies. We vary $k$ from 0 to 300.
\end{itemize}

\begin{figure}[t]
    \centering
    \begin{tabular}{c}
         \includegraphics[width=0.8\textwidth]{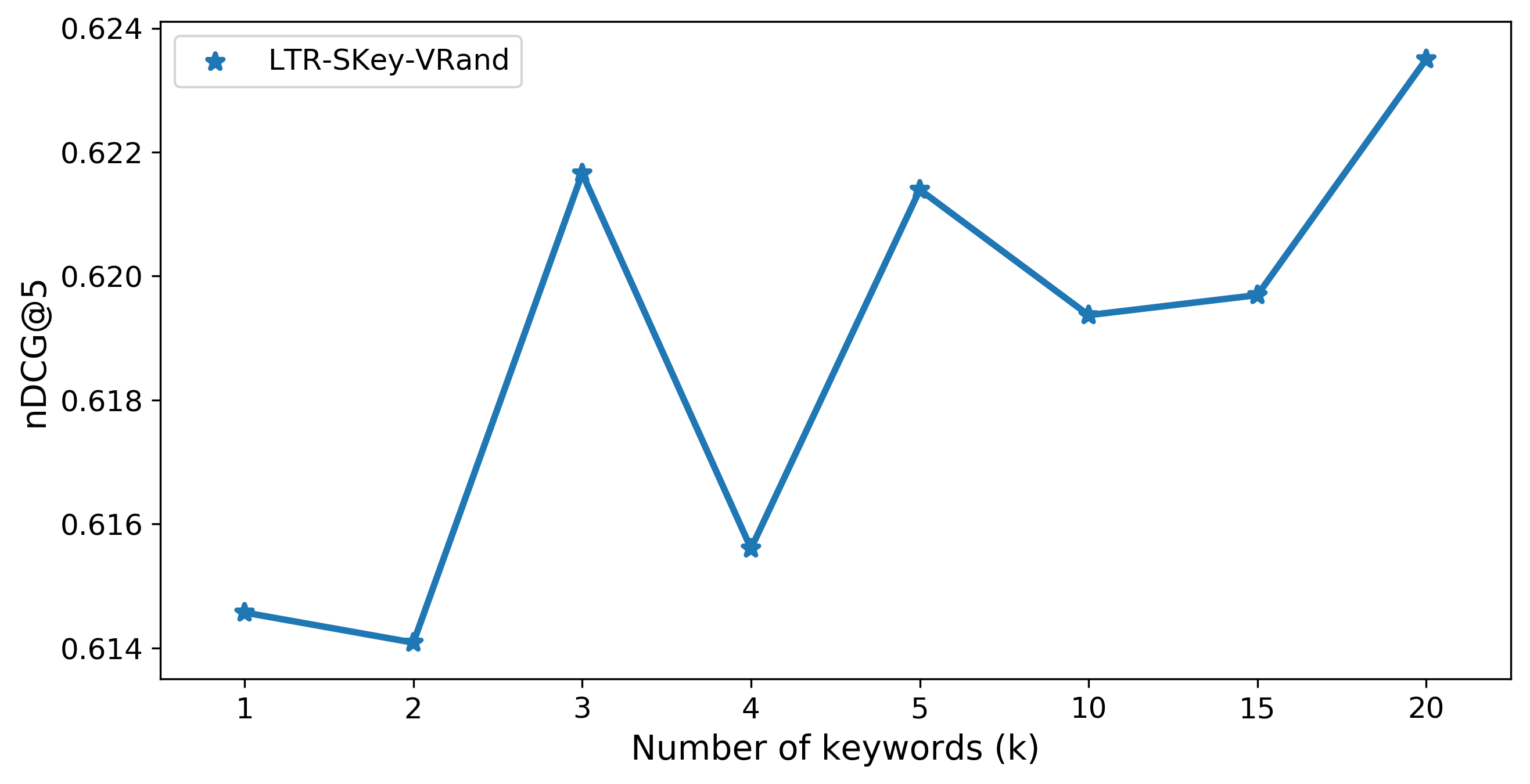} \\
         \includegraphics[width=0.8\textwidth]{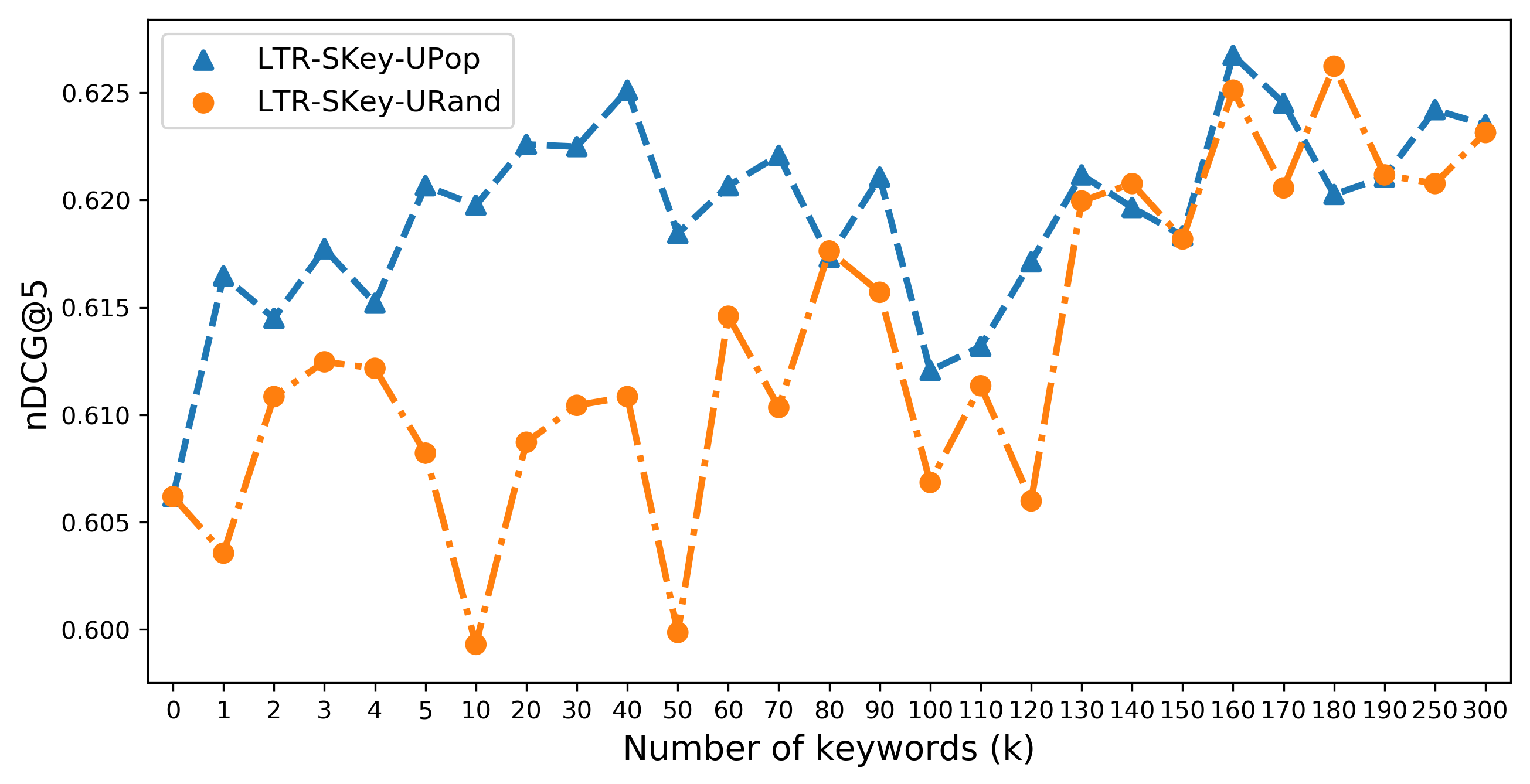} 
    \end{tabular}
    \caption{Performance of LTR-S in terms of nDCG@5 using different number of keywords ($k$).}
    \label{fig:keywords}
\end{figure}

As we can see in Figure~\ref{fig:keywords}, the performance of LTR-SKey-VRand increases as we increase the number of randomly selected keywords per venue. LTR-SKey-URand, on the other hand, shows a different behavior. We see that while in general having more keywords in the user's profile benefits the model, selecting $k$ keywords from the profile in a random order results in an inconsistent behavior of the model. For example, we observe that even in some cases (e.g., $k=10$) the performance of the model is lower than a model trained with no keywords. LTR-SKey-UPop behaves differently and, generally, its performance improves as we increase $k$. This shows that the popularity of a keyword in a user's profile is a good indicator of its importance to the user. We also see that the best performance is achieved when $k=160$, suggesting that applying a dimensionality reduction on the keywords space can help us model the users more effectively, something that we studied in \cite{DBLP:conf/ecir/Aliannejadi17}.
\section{Conclusions and Future Work}\label{Conclusions}
In this paper we proposed a set of similarity scores for recommending venues based on content- and social-based information.
As content, we used a frequency-based strategy to model venue categories. Social-centric scores consisted of online reviews on LBSNs and keywords that are automatically extracted from online reviews. We modeled the reviews using a classifier per user and used the same frequency-based strategy to model the keywords.
Experimental results corroborated the effectiveness of our approach and showed that combining social-centric scores outperforms all other scores combinations, as well as the baselines. Moreover, we studied the impact of the number of reviews and keywords per venue on the system's performance. Our results showed that selecting a certain number of reviews based on their timestamp or author's activity improves a system's efficiency and effectiveness. Also, selecting the $k$ most repeated keywords in a user's profile improves the efficiency of our model, indicating that reducing the dimensionality of  venue taste keywords in a smarter way can be beneficial, something that we explored in~\cite{DBLP:conf/ecir/Aliannejadi17}.

In the future, we plan to explore other keyword modeling approaches such as average word embedding, which has been proven to be effective in other domains~\cite{DBLP:conf/cikm/AliannejadiZCC18,DBLP:conf/sigir/AliannejadiZCC18}. Also, the availability of a massive number of online reviews has motivated us to leverage them to perform semi-supervised learning of the review classifier~\cite{DBLP:conf/acl-alta/AliannejadiKKG14,DBLP:conf/nips/BennettD98}.

\section*{Acknowledgment}
This work was partially supported by the Swiss National Science Foundation (SNSF) under the project ``Relevance Criteria Combination for Mobile IR (RelMobIR).''

\bibliographystyle{splncs03}
\bibliography{main} 

\begin{thebibliography}{10}
\providecommand{\url}[1]{\texttt{#1}}
\providecommand{\urlprefix}{URL }

\bibitem{alian2015}
Aliannejadi, M., Bahrainian, S.A., Giachanou, A., Crestani, F.: University of
  {L}ugano at {TREC} 2015: {C}ontextual {S}uggestion and {T}emporal
  {S}ummarization tracks. In: {TREC} 2015. {NIST} (2015)

\bibitem{alianSigir17}
Aliannejadi, M., Crestani, F.: Venue appropriateness prediction for
  personalized context-aware venue suggestion. In: {SIGIR} 2017. pp.
  1177--1180. {ACM} (2017)

\bibitem{DBLP:conf/iir/AliannejadiC18}
Aliannejadi, M., Crestani, F.: A collaborative ranking model with contextual
  similarities for venue suggestion. In: {IIR}. {CEUR} Workshop Proceedings,
  vol. 2140. CEUR-WS.org (2018)

\bibitem{DBLP:journals/tois/AliannejadiC18}
Aliannejadi, M., Crestani, F.: Personalized context-aware point of interest
  recommendation. {ACM} Trans. Inf. Syst.  36(4),  45:1--45:28 (2018)

\bibitem{DBLP:conf/acl-alta/AliannejadiKKG14}
Aliannejadi, M., Kiaeeha, M., Khadivi, S., Ghidary, S.S.: Graph-based
  semi-supervised conditional random fields for spoken language understanding
  using unaligned data. In: {ALTA} 2014. pp. 98--103. {ACL} (2014)

\bibitem{DBLP:conf/airs/AliannejadiMC16}
Aliannejadi, M., Mele, I., Crestani, F.: User model enrichment for venue
  recommendation. In: {AIRS} 2016. pp. 212--223. Springer (2016)

\bibitem{alianTREC2016}
Aliannejadi, M., Mele, I., Crestani, F.: Venue appropriateness prediction for
  contextual suggestion. In: {TREC} 2016. {NIST} (2016)

\bibitem{alianSigir17collection}
Aliannejadi, M., Mele, I., Crestani, F.: A cross-platform collection for
  contextual suggestion. In: {SIGIR} 2017. pp. 1269--1272. {ACM} (2017)

\bibitem{DBLP:conf/sac/AliannejadiMC17}
Aliannejadi, M., Mele, I., Crestani, F.: Personalized ranking for context-aware
  venue suggestion. In: {SAC} 2017. pp. 960--962. {ACM} (2017)

\bibitem{DBLP:conf/ecir/Aliannejadi17}
Aliannejadi, M., Rafailidis, D., Crestani, F.: Personalized keyword boosting
  for venue suggestion based on multiple {LBSNs}. In: {ECIR} 2017. pp.
  291--303. Springer (2017)

\bibitem{DBLP:conf/ictir/AliannejadiRC18}
Aliannejadi, M., Rafailidis, D., Crestani, F.: A collaborative ranking model
  with multiple location-based similarities for venue suggestion. In: {ICTIR}.
  pp. 19--26. {ACM} (2018)

\bibitem{8661539}
{Aliannejadi}, M., {Rafailidis}, D., {Crestani}, F.: A joint two-phase
  time-sensitive regularized collaborative ranking model for point of interest
  recommendation. IEEE Transactions on Knowledge and Data Engineering  (2019)

\bibitem{DBLP:conf/cikm/AliannejadiZCC18}
Aliannejadi, M., Zamani, H., Crestani, F., Croft, W.B.: In situ and
  context-aware target apps selection for unified mobile search. In: {CIKM}.
  pp. 1383--1392. {ACM} (2018)

\bibitem{DBLP:conf/sigir/AliannejadiZCC18}
Aliannejadi, M., Zamani, H., Crestani, F., Croft, W.B.: Target apps selection:
  Towards a unified search framework for mobile devices. In: {SIGIR}. pp.
  215--224. {ACM} (2018)

\bibitem{DBLP:conf/amt/BahrainianBSD10}
Bahrainian, S.A., Bahrainian, S.M., Salarinasab, M., Dengel, A.: Implementation
  of an intelligent product recommender system in an e-store. In: {AMT}. pp.
  174--182. Springer (2010)

\bibitem{DBLP:conf/nips/BennettD98}
Bennett, K.P., Demiriz, A.: Semi-supervised support vector machines. In:
  {NIPS}. pp. 368--374. The {MIT} Press (1998)

\bibitem{DBLP:conf/uai/BreeseHK98}
Breese, J.S., Heckerman, D., Kadie, C.M.: Empirical analysis of predictive
  algorithms for collaborative filtering. In: {UAI} 1998. pp. 43--52. Morgan
  Kaufmann (1998)

\bibitem{DBLP:journals/umuai/ChenCW15}
Chen, L., Chen, G., Wang, F.: Recommender systems based on user reviews: the
  state of the art. User Modeling and User-Adapted Interaction  25(2),  99--154
  (2015)

\bibitem{DBLP:conf/cikm/FerenceYL13}
Ference, G., Ye, M., Lee, W.: Location recommendation for out-of-town users in
  location-based social networks. In: {CIKM} 2013. pp. 721--726. {ACM} (2013)

\bibitem{DBLP:conf/recsys/GaoTHL13}
Gao, H., Tang, J., Hu, X., Liu, H.: Exploring temporal effects for location
  recommendation on location-based social networks. In: {RecSys} 2013. pp.
  93--100. {ACM} (2013)

\bibitem{DBLP:journals/csur/GiachanouC16}
Giachanou, A., Crestani, F.: Like it or not: {A} survey of twitter sentiment
  analysis methods. {ACM} Comput. Surv.  49(2),  28:1--28:41 (2016)

\bibitem{DBLP:journals/cacm/GoldbergNOT92}
Goldberg, D., Nichols, D.A., Oki, B.M., Terry, D.B.: Using collaborative
  filtering to weave an information tapestry. Commun. {ACM}  35(12),  61--70
  (1992)

\bibitem{DBLP:conf/recsys/GriesnerAN15}
Griesner, J., Abdessalem, T., Naacke, H.: {POI} recommendation: Towards fused
  matrix factorization with geographical and temporal influences. In: {RecSys}
  2015. pp. 301--304. {ACM} (2015)

\bibitem{DBLP:conf/recsys/KaratzoglouABO10}
Karatzoglou, A., Amatriain, X., Baltrunas, L., Oliver, N.: Multiverse
  recommendation: n-dimensional tensor factorization for context-aware
  collaborative filtering. In: RecSys 2010. pp. 79--86. {ACM} (2010)

\bibitem{koren2008factorization}
Koren, Y.: Factorization meets the neighborhood: a multifaceted collaborative
  filtering model. In: {SIGKDD} 2008. pp. 426--434. {ACM} (2008)

\bibitem{DBLP:journals/tois/LiJHL17}
Li, X., Jiang, M., Hong, H., Liao, L.: A time-aware personalized
  point-of-interest recommendation via high-order tensor factorization. {ACM}
  Trans. Inf. Syst.  35(4),  31:1--31:23 (2017)

\bibitem{DBLP:journals/ftir/Liu09}
Liu, T.: Learning to rank for information retrieval. Foundations and Trends in
  Information Retrieval  3(3),  225--331 (2009)

\bibitem{DBLP:conf/cikm/RafailidisC16}
Rafailidis, D., Crestani, F.: Joint collaborative ranking with social
  relationships in top-n recommendation. In: {CIKM} 2016. pp. 1393--1402. {ACM}
  (2016)

\bibitem{DBLP:conf/www/SarwarKKR01}
Sarwar, B.M., Karypis, G., Konstan, J.A., Riedl, J.: Item-based collaborative
  filtering recommendation algorithms. In: {WWW} 2001. pp. 285--295. {ACM}
  (2001)

\bibitem{DBLP:conf/trec/Yang015}
Yang, P., Fang, H.: University of delaware at {TREC} 2015: Combining opinion
  profile modeling with complex context filtering for contextual suggestion.
  In: {TREC} 2015. {NIST} (2015)

\bibitem{yang2015opinions}
Yang, P., Wang, H., Fang, H., Cai, D.: Opinions matter: a general approach to
  user profile modeling for contextual suggestion. Information Retrieval
  Journal  18(6),  586--610 (2015)

\bibitem{DBLP:conf/sigir/YeYLL11}
Ye, M., Yin, P., Lee, W., Lee, D.L.: Exploiting geographical influence for
  collaborative point-of-interest recommendation. In: {SIGIR} 2011. pp.
  325--334. {ACM} (2011)

\bibitem{DBLP:journals/tois/YinCSHC14}
Yin, H., Cui, B., Sun, Y., Hu, Z., Chen, L.: {LCARS:} {A} spatial item
  recommender system. {ACM} {TOIS}  32(3),  11:1--11:37 (2014)

\bibitem{DBLP:conf/ictai/YuanJGCYA16}
Yuan, F., Jose, J.M., Guo, G., Chen, L., Yu, H., Alkhawaldeh, R.S.: Joint
  geo-spatial preference and pairwise ranking for point-of-interest
  recommendation. In: {ICTAI} 2016. pp. 46--53. {IEEE} (2016)

\bibitem{DBLP:conf/sigir/YuanCMSM13}
Yuan, Q., Cong, G., Ma, Z., Sun, A., Magnenat{-}Thalmann, N.: Time-aware
  point-of-interest recommendation. In: {SIGIR} 2013. pp. 363--372. {ACM}
  (2013)

\bibitem{DBLP:conf/cikm/YuanCS14}
Yuan, Q., Cong, G., Sun, A.: Graph-based point-of-interest recommendation with
  geographical and temporal influences. In: {CIKM} 2014. pp. 659--668. {ACM}
  (2014)

\bibitem{DBLP:journals/tois/ZhangLW16}
Zhang, C., Liang, H., Wang, K.: Trip recommendation meets real-world
  constraints: {POI} availability, diversity, and traveling time uncertainty.
  {ACM} Trans. Inf. Syst.  35(1),  5:1--5:28 (2016)

\bibitem{DBLP:conf/sigir/ZhangC15}
Zhang, J., Chow, C.: Geosoca: Exploiting geographical, social and categorical
  correlations for point-of-interest recommendations. In: {SIGIR} 2015. pp.
  443--452. {ACM} (2015)

\bibitem{DBLP:journals/tist/ZhangDCLZ13}
Zhang, W., Ding, G., Chen, L., Li, C., Zhang, C.: Generating virtual ratings
  from chinese reviews to augment online recommendations. {ACM} {TIST}  4(1),
  9:1--9:17 (2013)

\end{thebibliography}

\end{document}